\documentclass[aps,prl,twocolumn,twoside,superscriptaddress,nofootinbib,preprintnumbers,floatfix,longbibliography]{revtex4-2}

\usepackage{hyperref}
\usepackage{graphicx}
\usepackage{amsmath}
\usepackage{mathrsfs}
\usepackage{xspace}
\usepackage{amsfonts}
\usepackage{color}
\usepackage{booktabs,siunitx}

\newcommand{\eq}[1]{Eq.~\eqref{eq:#1}}

\newcommand{\fig}[1]{Fig.~\ref{fig:#1}}

\newcommand{\nn}{\nonumber}

\allowdisplaybreaks[2]

\usepackage{xcolor}
\usepackage{marginnote}
\usepackage[normalem]{ulem}
\usepackage{dsfont,mathrsfs}

\begin{document}
\title{Heavy Quark Pair Energy Correlators:\\
From Profiling Partonic Splittings to Probing Heavy-Flavor Fragmentation}

\author{Jo\~{a}o Barata}
\email{joao.lourenco.henriques.barata@cern.ch}
\affiliation{CERN, Theoretical Physics Department, CH-1211, Geneva 23, Switzerland}

\author{Jasmine Brewer}
\email{jasmine.brewer@physics.ox.ac.uk}
\affiliation{Rudolf Peierls Centre for Theoretical Physics, University of Oxford, Oxford OX1 3PU, UK}

\author{Kyle Lee}
\email{kylel@mit.edu}
\affiliation{Center for Theoretical Physics -- a Leinweber Institute, Massachusetts Institute of Technology, Cambridge, MA 02139, USA}

\author{Jo\~{a}o M. Silva}
\email{joao.m.da.silva@tecnico.ulisboa.pt}
\affiliation{Departamento de Física Teórica y del Cosmos, Universidad de Granada, Campus de Fuentenueva, E-18071 Granada, Spain}
\affiliation{Laboratório de Instrumentação e Física Experimental de Partículas (LIP), Av. Prof. Gama Pinto, 2, 1649-003 Lisbon, Portugal}
\affiliation{Departamento de Física, Instituto Superior Técnico (IST), Universidade de Lisboa, Av. Rovisco Pais 1, 1049-001 Lisbon, Portugal}

\begin{abstract}
We introduce heavy-flavor energy correlators, $\langle \Psi |\mathcal{E}_{\mathcal{H}}(\vec{n}_1) \mathcal{E}_{\mathcal{H}}(\vec{n}_2)|\Psi \rangle$, as a powerful observable for profiling partonic splittings and characterizing heavy-flavor fragmentation. We present its collinear factorization, perform resummation, and demonstrate the angular distribution's sensitivity to both the heavy-quark pair splitting and their subsequent fragmentation. We then apply the heavy-flavor EEC to probe medium-induced effects, revealing its sensitivity to medium modifications to heavy-quark pair splitting functions and to the medium's spatial structure.
\end{abstract}

\preprint{CERN-TH-2025-172}
\preprint{MIT-CTP 5888}
\maketitle

\emph{\textbf{Introduction.}} High energy collider experiments provide unique access to probe the Standard Model through asymptotic energy detectors. In recent years, correlations of these energy flows in jet substructure have allowed to probe anomalous scaling~\cite{Lee:2022uwt,Komiske:2022enw,Chen:2020vvp}, transport properties and in-medium dynamics~\cite{Andres:2022ovj,Devereaux:2023vjz,Andres:2023ymw,Barata:2023bhh,Yang:2023dwc,Bossi:2024qho,Singh:2024vwb,Fu:2024pic,Barata:2024wsu,Apolinario:2025vtx}, the hadronization transition~\cite{Lee:2025okn,Chang:2025kgq,Kang:2025zto,Lee:2023npz,Lee:2023tkr,Chen:2024nfl,Barata:2024nqo, Herrmann:2025fqy}, and nuclear structure~\cite{Liu:2022wop,Liu:2024kqt,Liu:2023aqb,Kang:2023big} of Quantum Chromodynamics (QCD). For
a recent review, see~\cite{Moult:2025nhu}. At the heart of these jet studies lies the sensitivity to the underlying partonic splitting functions describing the universal behavior of scattering amplitudes in the collinear limit.

\begin{figure}[!t]
\begin{center}
\includegraphics[width=0.49\textwidth]{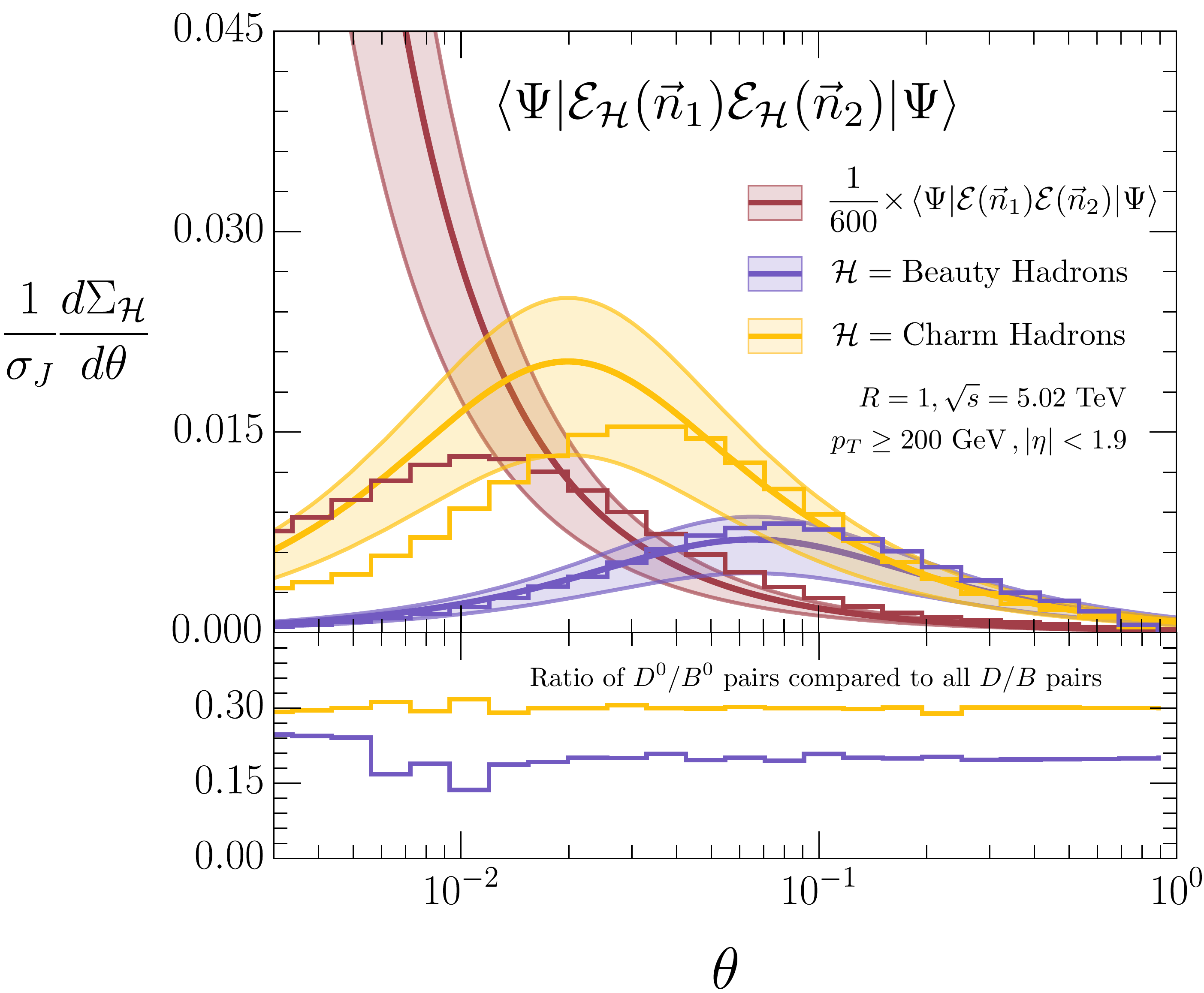}
\end{center}
\caption{Heavy‐flavor EEC on charm and bottom hadrons vs. inclusive EEC (rescaled by 1/600) for jets with $p_T\ge200\,$GeV. The heavy-flavor correlator isolates the splitting function $P_{g\to Q\bar{Q}}^{(1)}$ within the jet. The lower panel shows that selecting heavy‐hadron subsets only rescales the normalization via HQET matrix elements. Histograms are from \textsc{Pythia} and curves are NLL predictions of~\eq{Fact} with scale variations.}
\label{fig:EEC_vac}
\end{figure}

One of QCD’s open problems is how such parton splitting functions are reshaped by the primordial quark–gluon plasma (QGP) and how those partons ultimately recombine into color-neutral hadrons. In this Letter we show that two‐point heavy‐flavor energy correlator (EEC)
\begin{align}
\label{eq:EECheavy}
&\langle \Psi | \mathcal{E}_{\mathcal{H}}(\vec n_1)\,\mathcal{E}_{\mathcal{H}}(\vec n_2)| \Psi\rangle 
\equiv \frac{1}{\sigma_J}\int \frac{d\sigma_J}{dp_{T} d\eta}\\
&\times \sum_{i_1,i_2 \in \mathcal{H}\subset J}\frac{p_{T,i_1} p_{T,i_2}}{p_T^2}\prod_{j=1}^2\delta(\eta_j-\eta_{i_j})\delta(\phi_j-\phi_{i_j})\nonumber
\end{align}
within a jet state\footnote{We denote differential jet cross section as $\frac{d\sigma_J}{dp_T d\eta}$ for a given transverse momentum $p_T$ and rapidity $\eta$. The dependence on the momenta of particles inside the jet is also differential but left implicit. The total jet cross section, $\sigma_J$, is obtained by integrating over the desired $p_T$ and $\eta$ bins.} $|\Psi\rangle = |\Psi(\eta,p_T)\rangle$ at high energy collider experiments provides unique opportunities to address these fundamental problems. Here $\mathcal{E}_{\mathcal{H}}(\vec n)$ measures the energy flux carrying heavy‐quark quantum number $\mathcal{H}$\footnote{We denote by $\mathcal{H}$ a set of heavy hadrons containing a heavy quark $Q$ or heavy anti-quark $\bar Q$ quantum number.} along the direction $\vec{n}_1=\vec{n}_1(\eta_1,\phi_1)$ parameterized in the rapidity–azimuth plane. The angular dependence of these correlators directly probes the mass-dependent partonic splitting functions $P_{i\to Q_1 Q_2+X}$, describing the collinear splitting of parton $i$ into a pair of heavy parton $Q_1 Q_2$ plus additional collinear particles that \eq{EECheavy} is inclusive over, and simultaneously captures the subsequent fragmentation of $Q_i$ to a heavy hadron. The leading order contribution of such splitting comes from 
\begin{align}
P_{g\to Q\bar{Q}}^{(1)}(x,m_Q) = T_F\left[1-\frac{2 x(1-x)}{1-\epsilon}+\frac{2 m_Q^2}{(1-\epsilon) s}\right]\,,
\end{align}
where $s=(p_Q+p_{\bar{Q}})^2$ and $x$ is the momentum fraction of the outgoing $Q$.

Our proposal is part of many fruitful efforts to study heavy-flavor dynamics and parton splitting function~\cite{Dhani:2025fbk,Xing:2024yrb,Dhani:2024gtx,Blok:2023ugf,Yang:2024awm,Caletti:2023spr,Andres:2023ymw,Chen:2024nfl,Cunqueiro:2022svx,Craft:2022kdo,Li:2021gjw,Cal:2021fla,Casalderrey-Solana:2019ubu,Caucal:2019uvr,Li:2017wwc,Lee:2019lge,Chang:2017gkt,Milhano:2017nzm,Mehtar-Tani:2016aco,KunnawalkamElayavalli:2017hxo,Chien:2016led} within jet substructure and/or in medium~\cite{Armesto:2003jh, Sievert:2019cwq,Sievert:2018imd,Zhang:2003wk,Arleo:2012rs,Djordjevic:2003zk}, and demonstrates marked improvements in its ability to precisely characterize the heavy-flavor fragmentation~\cite{Czakon:2024tjr,vonKuk:2023jfd,Czakon:2022pyz,Kang:2016ofv,Bjorken:1977md,Suzuki:1977km,Peterson:1982ak,Mele:1990cw,Mele:1990yq,Jaffe:1993ie,Cacciari:2001cw,Cacciari:2002xb,Gardi:2005yi,Melnikov:2004bm,Cacciari:2005uk,Mitov:2006xs,Aglietti:2006yf,Fickinger:2016rfd,Bauer:2013bza,Dai:2023rvd,Yao:2018zze} rate and unambiguously tag partonic splittings involving heavy quark pair, all while sidestepping the infrared ambiguities that challenge traditional heavy-flavor jet definitions~\cite{Banfi:2006hf}.\footnote{See~\cite{Gauld:2022lem,Caola:2023wpj,Czakon:2022wam,Caletti:2022glq,Caletti:2022hnc,Fedkevych:2022mid,Behring:2025ilo} for recent proposals to resolve this.} The precise understanding of the heavy-flavor EEC in vacuum given in~\eq{EECheavy} gives us the means to use them as a handle to probe the nontrivial medium-induced modifications of splitting and even the geometrical shape of the medium.

\emph{\textbf{Heavy Flavor Energy Correlators in Vacuum.}}
Focusing on the longitudinal boost-invariant angle between the two detectors $\theta \equiv \sqrt{(\Delta \eta_{12})^2 + (\Delta \phi_{12})^2 }$, we first focus on the distribution
\begin{align}
\frac{1}{\sigma_J}\frac{d\Sigma_{\mathcal{H}}}{d\theta} \equiv& \int \prod_{i=1}^2d\phi_i d\eta_i \delta\Bigl(\theta - \sqrt{(\Delta \eta_{12})^2 + (\Delta \phi_{12})^2 }\Bigr)\nn\\
&\times\langle \Psi | \mathcal{E}_{\mathcal{H}}(\vec n_1)\,\mathcal{E}_{\mathcal{H}}(\vec n_2)| \Psi\rangle\,.
\end{align}
The cumulant of the collinear limit of the energy correlators for heavy-flavor hadrons distribution $z = \theta^2/4$ then factorizes as
\begin{align}
\label{eq:Fact}
\Sigma_{\mathcal{H}}\left(z, \frac{Q^2}{\mu^2}, \mu\right) &=\int_0^{z} dz^{\prime} \frac{1}{\sigma_J}\frac{d\Sigma_{\mathcal{H}}}{dz'} \\
&\hspace{-2cm}=\int_0^1 d x\, x^2 \vec{J}_{\mathcal{H}}\left(\ln \frac{z Q^2}{\mu^2}, m_Q,\mu\right) \cdot \vec{H}\left(x, \frac{Q^2}{\mu^2}, \mu\right)\,,\nonumber
\end{align}
where the subscript $\mathcal{H}$ indicates restriction of the correlator to heavy hadrons carrying the heavy-quark quantum number. Collinear factorization for the EEC in lepton colliders or jet substructure has been derived in~\cite{Dixon:2019uzg,Lee:2022ige,Lee:2024icn,Lee:2025okn}, and extended to charged-hadron measurements in~\cite{Jaarsma:2023ell,Lee:2023npz,Lee:2023tkr}. Our heavy-flavor EEC factorization parallels the charged-hadron case, replacing electric charge with heavy-quark quantum number.

Here, the hard and jet functions, denoted as $\vec{H}$ and $\vec{J}_{\mathcal{H}}$ respectively, are vectors in flavor space. The hard function $\vec{H}$ also depends on parton distribution functions and encodes the short-distance production of the fragmenting parton that initiates the jet state $| \Psi \rangle$. The jet function $\vec J_{\mathcal{H}}$ describes the angular correlation of two heavy hadrons inside the jet, and is therefore sensitive both to the angular scale $2p_T\sqrt{z}$ and to the heavy-quark mass $m_Q$. In the region $m_Q \ll 2p_T \sqrt{z} \ll p_T$, the mass dependence is power-suppressed, and the factorization reduces to the massless result. The renormalization group (RG) evolution between the  jet scale $\mu_J \sim 2p_T \sqrt{z}$ and the hard scale $\mu_H\sim p_T$ gives the resummation of collinear logarithms $\ln z$ consistent with the anomalous scaling predicted from the light-ray operator product expansion (OPE)~\cite{Chang:2020qpj,Kologlu:2019mfz,Hofman:2008ar,Chen:2021gdk,Chen:2023zzh}. On the other hand, in the region $m_Q \sim 2p_T \sqrt{z} \ll p_T$, the resummation only affects the normalization of the fixed order contribution to the EEC.

We will present the details of our factorization in the accompanying paper~\cite{Barata:2025xxx}.

In~\fig{EEC_vac}, we see that both our factorization predictions and Pythia distribution, as functions of $\theta$, are highly sensitive to the shape of the massive splitting $P_{g\to Q\bar{Q}}^{(1)}$, the leading splitting for producing two heavy flavors in the final-state. We observe its characteristic peak structure near $2p_T \sqrt{z}\sim m_Q$.  This is done without the need for defining flavor sensitive definitions of jet algorithms~\cite{Behring:2025ilo,Gauld:2022lem,Caola:2023wpj,Czakon:2022wam,Caletti:2022glq,Caletti:2022hnc,Fedkevych:2022mid}, but rather looking at the heavy flavor-heavy flavor correlations inclusively over inclusive jet state $| \Psi \rangle$ characterized only by $p_T$ and $\eta$. It is the heavy‐flavor EEC itself --- rather than an explicit jet‐flavor selection --- that singles out the $P_{g\to Q\bar Q}^{(1)}$ splitting at leading order, with sensitivity to other splittings (such as subsequent $Q\to Qg$ splittings of the produced $Q$ or $\bar{Q}$) also incorporated through resummation, which affects the normalization. Indeed, when compared with the inclusive measurement over all energy flux within the jet, $\langle \Psi|\mathcal{E}(\vec{n}_1) \mathcal{E}(\vec{n}_2) |\Psi\rangle$, with identical jet kinematics, we observe a striking difference.\footnote{Note that the inclusive jet calculation here is purely based on the OPE region. In the very small-angle region, one needs input from dihadron fragmentation functions~\cite{Lee:2025okn}.} This is unsurprising, as the inclusive structure is sensitive to an ensemble average over many different splitting functions.

\emph{\textbf{Heavy Flavor Fragmentation and Asymptotic Detectors.}}
Since we are considering correlations involving heavy flavor hadrons, the fixed order jet function that we evolve are sensitive to the fragmentation of heavy (anti-)quark into heavy-flavor hadrons. Due to heavy quark mass scale being perturbative, the heavy flavor fragmentation function\footnote{In general, correlations of asymptotic detectors involve heavy flavor track functions~\cite{Li:2021zcf,Chang:2013rca,Lee:2023xzv,Lee:2023tkr}, or more generally, detector functions~\cite{Gonzalez:2025xxx}. The detailed formalism of the heavy flavor track functions will be discussed in the accompanying long paper~\cite{Barata:2025xxx}.} $D_{Q\to H}(z,m_Q,\mu)$ can be perturbatively matched at the $\mu\sim m_Q$ scale onto the nonperturbative matrix element, $\chi_Q^{H}$, of (boosted) Heavy-Quark Effective Theory~\cite{Eichten:1989zv, Isgur:1989vq, Isgur:1990yhj, Grinstein:1990mj, Georgi:1990um,Korner:1991kf,Mannel:1991mc,
Fleming:2007qr, Fleming:2007xt}, describing the fragmentation rate to different heavy flavor hadron $H$ at the hadronization scale $\Lambda_{\rm QCD}$. Therefore, our heavy flavor correlations in the $Q\sqrt{z}\sim m_Q$ region become proportional to $\sum_{H,\bar{H}\in \mathcal{H}} \chi_Q^{H}\chi_{\bar{Q}}^{\bar{H}}$, which is equal to identity by heavy quark symmetry if $\mathcal{H}$ includes all the heavy flavor hadrons with quantum number $Q$ or $\bar{Q}$, rendering the calculation completely perturbative up to power corrections in $\Lambda_{\rm QCD}/m_Q$. 

On the other hand, even if we assume $\mathcal{H}$ includes only a subset of heavy flavor hadron states, the factorization tells us that the results are modified simply by an overall $\sum_{H,\bar{H}\in \mathcal{H}} \chi_Q^{H}\chi_{\bar{Q}}^{\bar{H}}$, which is simply a number describing the production fractions of heavy quark $Q$ and $\bar{Q}$ into a set of heavy-flavor hadrons in $\mathcal{H}$. In \fig{EEC_vac}, we consider taking $\mathcal{H} =  \{B^0,\bar{B}^0\}$ and $\mathcal{H} = \{D^0,\bar{D}^0\}$, respectively, for $Q=b$ and $Q=c$ in our Pythia simulations. With respect to all $B$ and $D$ hadron case, we observe that the shape is unmodified as expected, while the overall normalization is reduced by roughly $0.3$ and $0.2$, respectively. The normalization change implies that production fraction for $B^0$ and $D^0$ are $\chi_b^{B^0} =\chi_{\bar{b}}^{\bar{B}^0} \approx \sqrt{0.2} \approx 0.45$ and  $\chi_c^{D^0} =\chi_{\bar{c}}^{\bar{D}^0} \approx \sqrt{0.3} \approx 0.55$, which is roughly consistent with the expectation from $B^0$ fractions $0.407\pm0.007$ measured at  $e^+e^-$ colliders~\cite{HFLAV:2016hnz,ALEPH:2001ccp} and $D^0$ fractions $0.544 \pm 0.022$ computed using data from HERA and $e^+e^-$ colliders~\cite{ZEUS:2013fws,Gladilin:1999pj}. The slight difference in the production rates of $B^0$ and $D^0$ arises from heavy-quark symmetry breaking, which scales as $\Lambda_{\rm QCD}/m_Q$.

Experimentally tagging heavy-flavor hadrons within boosted jets is highly challenging, and analyses involving double heavy-flavor tagging are even more scarce~\cite{ATLAS:2018zhf,CDF:2004mmv}. However, our result demonstrates that even with only a subset of heavy flavor hadrons accessed (whether it be restricted by particular flavor or their particular branching channel), the overall shape of our results will remain unmodified, while the shift in normalization will tell us the fraction of heavy hadron pairs effectively tagged within the jet. Note also that even if tagging capabilities are nontrivial functions of energies of the heavy hadron, rather than simply restricted by flavors, this can be accounted for as energy distribution is perturbatively computable.

\begin{figure}[t!]
\begin{center}
\includegraphics[width=0.46\textwidth]{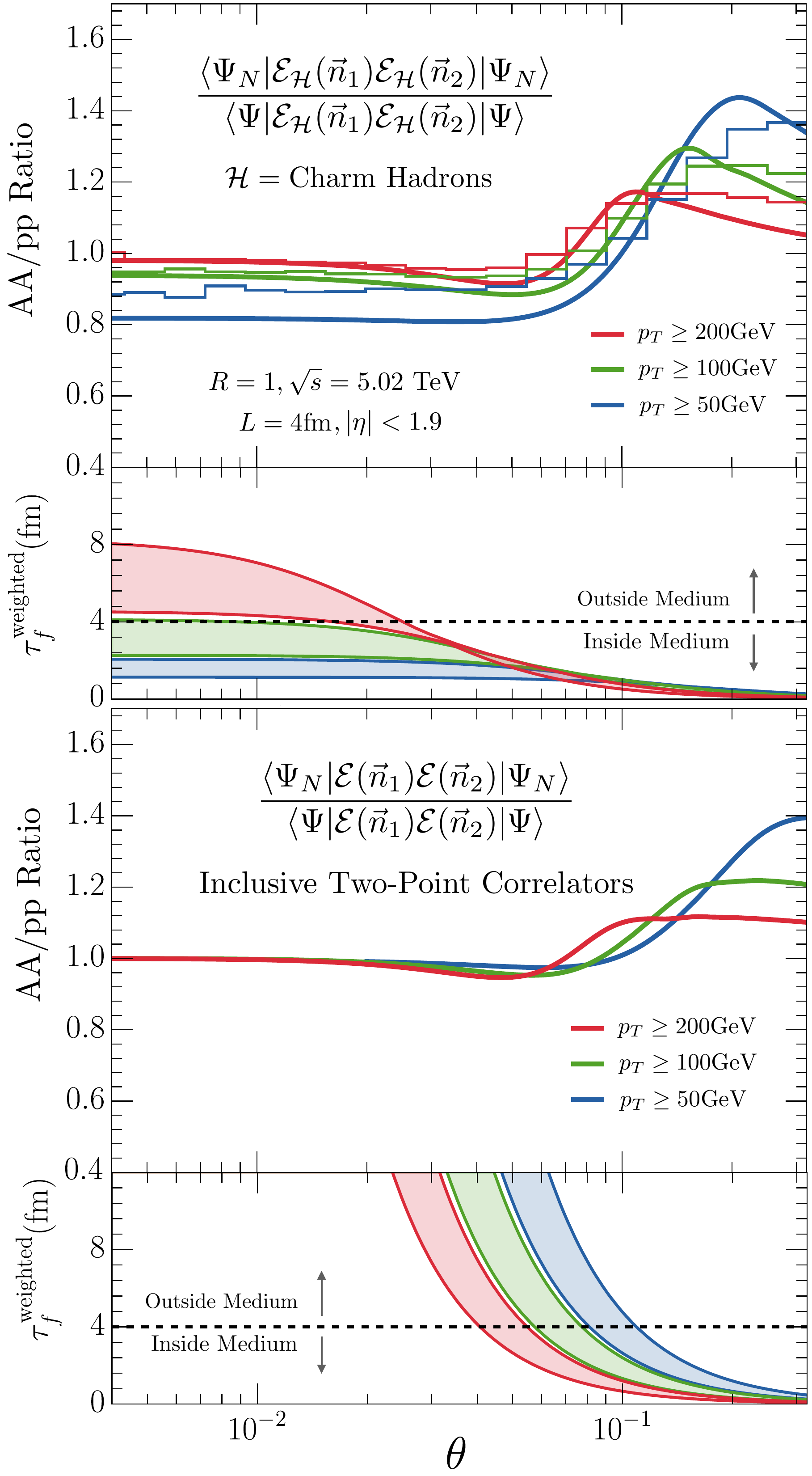}
\end{center}
\caption{Medium modification of heavy-flavor and inclusive EEC in a medium of length $L = 4~$fm for several jet $p_T$ bins. Heavy-flavor histograms (PYTHIA) use a reweighted $g\to c\bar{c}$ kernel and solid curves are NLL predictions of~\eq{Fact} with medium-modified splitting functions. Lower panels show $\tau_f^{\rm weighted}$ given in~\eq{tauweighted} for light- and charm-quark pairs, with uncertainty from $0.55\,p_T \leq E_g \leq p_T$. The medium-to-vacuum ratio deviates from unity occur where $\tau_f^{\rm weighted}/L$ is small.}
\label{fig:EEC_med}
\end{figure}

\emph{\textbf{Heavy Flavor Energy Correlators in Medium.}}
With precise control on effectively tagging the underlying splitting function $P^{(1)}_{g\to Q\bar{Q}}$ in vacuum, the in-medium heavy-flavor EEC provides an unambiguous determination of its modification, eliminating sensitivity to flavor mixing.

\begin{figure*}[!t] 
    \centering
\includegraphics[width=0.78\textwidth]{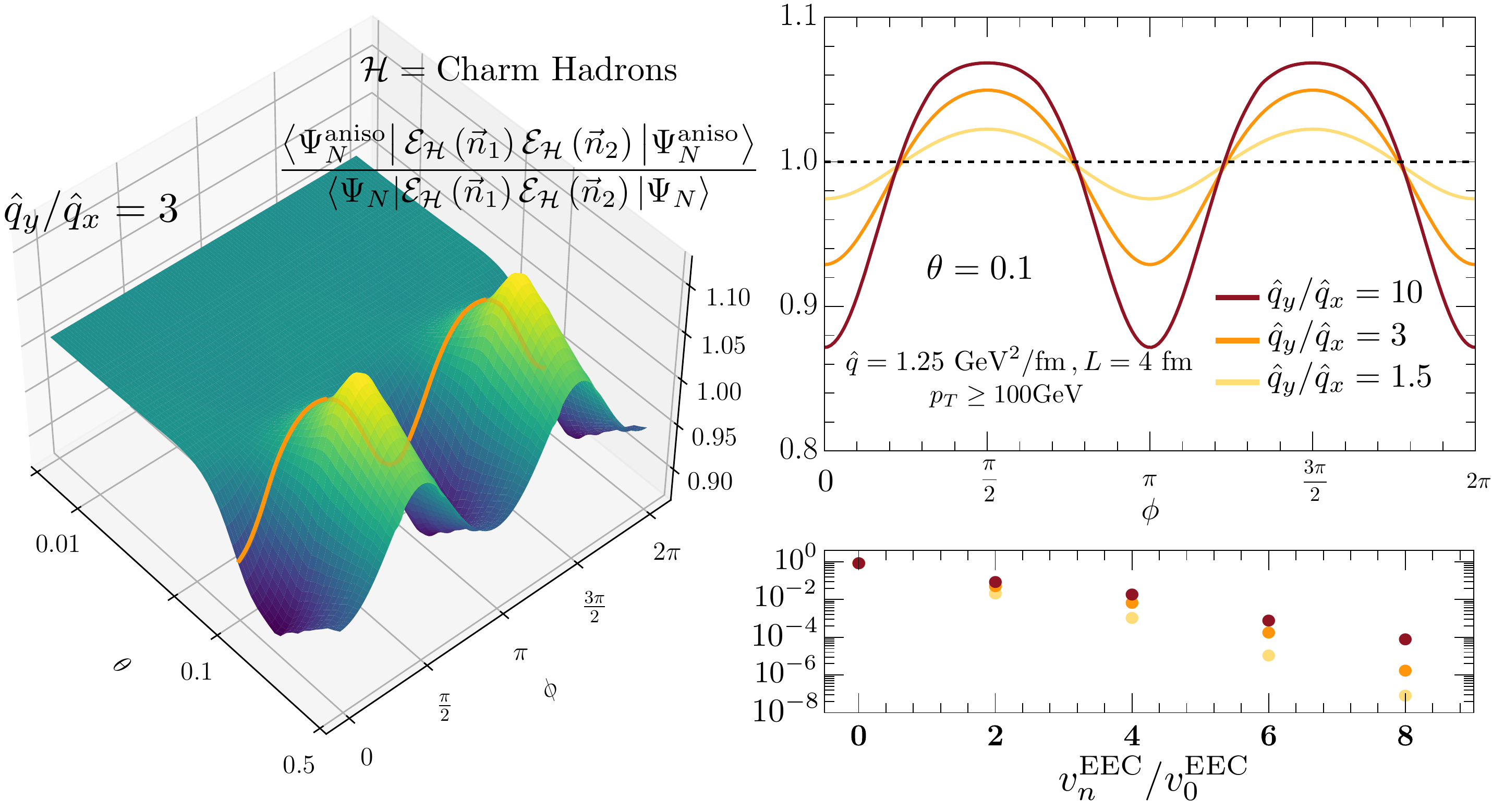}
\caption{Ratio of the heavy‐flavor EEC inside a jet produced in an anisotropic medium to that in an isotropic medium. Anisotropy induces a nontrivial azimuthal dependence in $\phi$, which can be expanded in even harmonics. For $\theta=0.1$, the right panel displays the $\phi$ dependence of this ratio for several magnitudes of anisotropy in the jet‐quenching parameter, along with its harmonic coefficients $v_{2n}^{\rm EEC}$ in
the ratio $\frac{\langle\Psi_N^{\rm aniso}|\mathcal{E}_{\mathcal{H}}(\vec{n}_1)\mathcal{E}_{\mathcal{H}}(\vec{n}_2)|\Psi_N^{\rm aniso}\rangle}{\langle\Psi_N|\mathcal{E}_{\mathcal{H}}(\vec{n}_1)\mathcal{E}_{\mathcal{H}}(\vec{n}_2)|\Psi_N\rangle}=\sum_{n=0}^\infty v_{2n}^{\rm EEC} \cos(2n\phi)$. As the anisotropy increases, higher‐order harmonics become more pronounced.}    \label{fig:anisotropy}
\end{figure*}

The medium modification $P^{\rm{med}(1)}_{g\to Q\bar{Q}}$ splitting was calculated in the 
limit of a dense medium in~\cite{Attems:2022ubu,Attems:2022otp}, in the large-$N_c$ approximation. It describes how interactions with the medium broaden the relative transverse momentum of the $Q\bar{Q}$ pair and enhance the overall $g\to Q\bar{Q}$ production rate. In this letter, we use a \textit{static brick approximation} for the medium with fixed length $L=4~$fm and a quenching parameter $\hat{q}=1.25~$GeV$^2$/fm. Under this approximation, the $P^{\rm{med}(1)}_{g\to Q\bar{Q}}$ splitting function depends on the splitting's formation time $\tau_f = 2 E_g / (p_Q+p_{\bar{Q}})^2$, which determines whether the pair is formed inside the medium, i.e. $\tau_f\lesssim L$.

In the collinear limit, the formation time takes the form
  $  \tau_f = \frac{2 x(1-x) E_g^{-1}}{m_Q^2/E_g^2 + x^2 (1-x)^2 \theta^2}$
where $E_g$ is the gluon energy. In the $\theta\to 0$ limit, the light-parton formation time diverges, while that of the heavy-flavor stays finite due to the $m_Q$ term. Consequently, at very small angles light-parton splittings occur outside the medium, whereas heavy-quark splittings can occur in-medium for any $\theta$ provided the $m_Q^2/E_g$ ratio is sufficiently large. This is consistent with the well-known observation that heavy-flavor hadrons form early in the medium, making them effective probes of medium properties.

In \fig{EEC_med}, we illustrate the modification of the EEC measured in medium computed based on~\eq{Fact} with the medium modified splitting in the jet function.\footnote{The medium-induced splittings, beyond quark pair production, were obtained retaining only the ``factorizable" contribution (see e.g.~\cite{Blaizot:2012fh,Apolinario:2014csa, Isaksen:2023nlr}).} For the inclusive EEC, where most of the underlying splittings are coming from light-partons, we observe that the EEC ratio between medium and vacuum goes to $1$ in the small angle and becomes enhanced at large angles. Motivated by the energy weighting in the EEC, we compare with the weighted formation time 
\begin{align}
\label{eq:tauweighted}
\tau_f^{\rm weighted}\equiv\int dx\, 2x\,(1-x)\,\tau_f\,P^{(1)}_{g\to Q\bar{Q}}(x,m_Q), 
\end{align}
and observe that this is consistent with roughly where the weighted formation time becomes larger than the medium length. On the other hand, we find that the weighted formation time for $c\bar{c}$ pair remains finite even in the small angle, both the analytic result and Pythia simulations\footnote{Here, analytic results include all partonic channels; in the Monte Carlo simulations we follow the re-weighting prescription in~\cite{Attems:2022otp}, which is justified as the heavy-flavor EEC is primarily sensitive to the $g\to Q\bar{Q}$ splitting.} show modification even in the small angle region with varying magnitude depending on the $m_Q/E_g$ ratio. For sufficiently large $p_T$ jets, the $c\bar{c}$ pair is largely formed outside the medium at small angle. In contrast, for small $p_T$ jets, there is no angular region where the pair can form outside the medium, even at small angles. This kinematic interplay makes the heavy-flavor EEC a particularly interesting observable for parameterizing medium modification effects across different kinematic regions.

In the large angle region, we find the enhancement characteristic of several EEC calculations~\cite{Andres:2022ovj,Barata:2023bhh,Yang:2023dwc} and measurements~\cite{CMS-PAS-HIN-23-004,talk_Jussi,talk_Anjali,talk_Ananya}. For inclusive jets, this enhancement is connected to both the higher-twist contributions and uncorrelated soft radiation~\cite{Andres:2024xvk,Barata:2025fzd,Barata:2024ukm}. We note that one advantage of the exclusive nature of the proposed heavy-flavor EEC is that the latter contributions are suppressed, since heavy flavors cannot be thermally produced at leading order. On the other hand, as we approach smaller angles, we observe a suppression due to momentum broadening, where in-medium small angles are suppressed as they are broadened. For the inclusive case, this suppression region is restricted to the small region of the phase space, whereas the heavy-flavor EEC is suppressed for the wider range of small angle regions formed inside the medium. Note that momentum is broadened to $k_\perp^2= x^2 (1-x)^2\theta^2 E_g^2 \sim \hat{q} L$ and thus angles parametrically smaller can be identified as the suppression region due to broadening, which is consistent with the different medium modification regions we find.

We find qualitatively similar results for the $b\bar{b}$ case, which, due to its large mass, is formed within the medium for all $p_T$ kinematics considered for $c\bar{c}$ and yet experiences smaller medium effects, such as reduced momentum broadening, because of dead-cone-suppressed radiation. Details of the $b\bar{b}$ case in medium will be discussed in a forthcoming paper~\cite{Barata:2025xxx}.

\emph{\textbf{Probing the Shape of the Medium.}}
Having explicitly identified the heavy pair splitting using the heavy EEC, we wish now to illustrate how this observable can be further utilized to construct a tomographic picture of the underlying QGP matter. To this end, we consider the case of anisotropic matter  modeled by an axis dependent jet quenching transport coefficient, i.e. $\hat{q}_x\neq\hat{q}_y$, with $\hat{q} = \frac{\hat{q}_x+\hat{q}_y}{2}$. Such anisotropy leads to enhanced momentum broadening along a given axis~\cite{Hauksson:2023tze}, directly modifying the fragmentation pattern as the in-medium splitting $P_{g\to Q\bar{Q}}^{\rm med}$ acquires an azimuthal $\phi$ dependence~\cite{Barata:2024bqp}, in close analogy to studies of jet evolution in the early stages of heavy-ion collisions or in hydrodynamically evolving fireballs~\cite{Barata:2023zqg,Ipp:2020mjc,Kuzmin:2023hko,Sadofyev:2021ohn,Armesto:2004pt,Barata:2024xwy,Avramescu:2024poa,Boguslavski:2023alu,Carrington:2021dvw,Du:2023izb,Apolinario:2017sob,Fu:2022idl}.

In~\fig{anisotropy}, we show the double differential heavy‐flavor EEC for a jet evolving in anisotropic medium $|\Psi_N^{\rm aniso}\rangle$, divided by that for a jet in an isotropic medium $|\Psi_N\rangle$. Relative to the isotropic case, at small $\theta$ we see a relatively flat ratio, while in the large angle region there is a non-trivial $\pi$-periodic azimuthal modulation of the EEC, resulting from the anisotropic structure of the medium. In particular, this $\phi$-modulation can be decomposed into an even harmonics series~\fig{anisotropy}, where the importance of larger harmonic increases directly with the size of the anisotropy. A direct experimental measurement of such structure would provide an unique new window into the early time dynamics of heavy ion collisions, which have so far not been directly probed.

\emph{\textbf{Conclusions and Outlook.}} In this Letter, we have shown that heavy-flavor energy–energy correlators cleanly isolate the heavy-quark pair splitting function and its fragmentation dynamics. In vacuum, we provided precise predictions for its characteristic angular distribution and normalization --- modulated by the choice of heavy-hadron species and suppressed by roughly three orders of magnitude compared to the inclusive EEC --- demonstrating its exceptional sensitivity to heavy-quark pair splittings within a jet.

When embedded in a static quark–gluon plasma, the heavy-flavor EEC exhibits a clear in-medium signature of modified heavy-quark pair splittings: a higher-twist–driven enhancement at large opening angles and a suppression at small angles from momentum broadening. The angular regions of these effects align with the weighted formation time $\tau_f^{\rm weighted}$ in~\eq{tauweighted} relative to the medium length $L$ and the transverse broadening scale.  Introducing anisotropic transport coefficients further induces measurable azimuthal asymmetries in the heavy-quark pair splitting, opening a novel tomographic window on the QGP’s shape. Finally, the heavy-flavor EEC is largely insensitive to soft medium response~\cite{Bossi:2024qho,Yang:2023dwc} because thermal heavy-quark production is negligible.

Looking ahead, experimental measurements of heavy-flavor EECs in heavy-ion collisions at high-energy colliders promise to sharpen our understanding of heavy-flavor fragmentation and medium-induced modifications to heavy-quark splitting with minimal sensitivity to competing effects. This enhanced precision will be crucial for studies of Higgs and top-quark production~\cite{ATLAS:2018kot,CMS:2018nsn,ATLAS:2018fwl,CMS:2020grm} where heavy-flavor dynamics enter both as signal and background. Moreover, our framework can be extended to other heavy-flavor correlator observables, such as ones involving quarkonium in jets~\cite{Bain:2017wvk,Kang:2017yde,Chen:2024nfl,Wang:2025drz,Copeland:2025osx} and/or in the QGP~\cite{Matsui:1986dk}, providing fresh insight into the longstanding puzzle of heavy quarkonium production~\cite{Brambilla:2010cs}.

\begin{acknowledgments}

{\it Acknowledgements.}---We thank Carlota Andrés, Philipp Aretz, Ibrahim Chahrour, Hao Chen, Fabio Dominguez, Terry Generet, Philip Harris, Andre Hoang, Jack Holguin, Matt LeBlanc, Ezra Lesser, Johannes Michel, Ian Moult, Matthew Nguyen, Rene Poncelet, Jennifer Roloff, Simon Rothman, Iain Stewart, Varun Vaidya, and  Nima Zardoshti for useful discussions. K.L. was supported by the U.S. Department of Energy, Office of Science, Office of Nuclear Physics from DE-SC0011090. J.M.S. has been supported by MCIN/AEI (10.13039/501100011033) and ERDF (grant PID2022-139466NB-C21) and by Consejería de Universidad, Investigación e Innovación, Gobierno de España and Unión Europea – NextGenerationEU under grant AST22\_6.5. JB acknowledges support through a University Research Fellowship from the Royal Society under grant URF\textbackslash R1\textbackslash 241231, from the Leverhulme Trust under grant LIP2020-014, and from the UK Science and Technology Facilities Council (STFC) under grant ST/T000864/1.
\end{acknowledgments}

\textbf{\textit{End Matter: Energy Loss and Nuclear PDFs.}}
In~\fig{EEC_med}, we included only the medium-modified splitting functions in the jet function appearing in~\eq{Fact}. The central goal of our work is to isolate the heavy-flavor splitting and study how it is modified in the medium. In reality, additional contributions arise from energy loss and from initial-state nuclear effects. In this End Matter, we discuss how these effects alter the results and how their impact can be mitigated to largely isolate the heavy-flavor splitting dynamics.

\begin{figure}
    \centering
\includegraphics[width=0.48\textwidth]{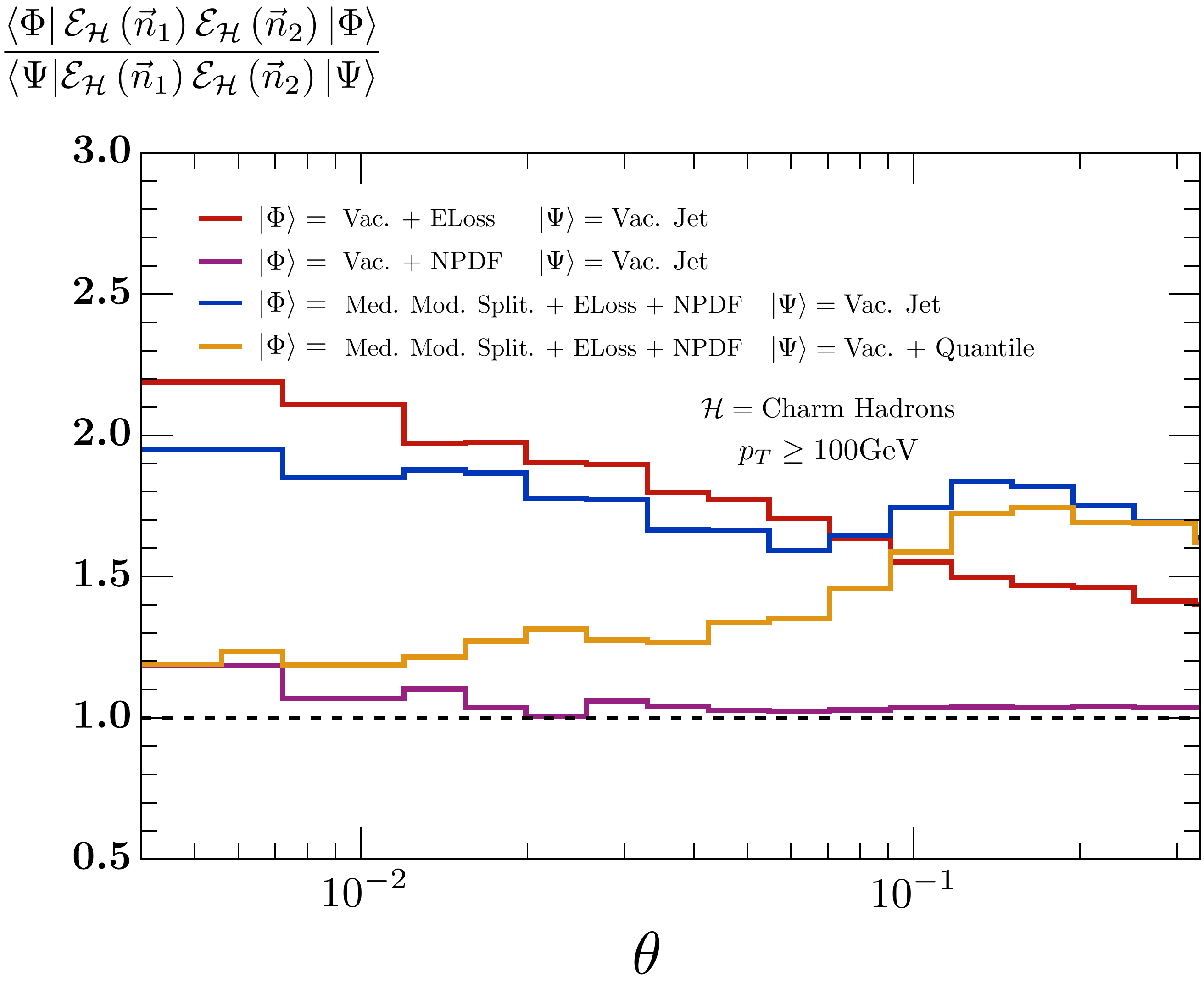}
\caption{Impact of energy loss, nPDFs, and medium-modified splittings on the heavy-flavor EEC. The quantile procedure removes the small-angle enhancement from $p_T$ migration, while the overall normalization remains enhanced due to the reduced energy loss of heavy quarks.}
\label{fig:EEC_eloss}
\end{figure}

We incorporate initial-state nuclear effects by employing nuclear parton distribution functions (nPDFs) in the hard function of~\eq{Fact}, using the EPPS16 set~\cite{Eskola:2016oht}. As shown in~\fig{EEC_eloss} by the purple curve, the heavy-flavor EEC with nPDFs is nearly identical to the result obtained with proton PDFs. This stems from the exclusive nature of the observable: the heavy-flavor EEC is sensitive only to jets containing at least two heavy-flavor hadrons, which are largely gluon jets. This contrasts with the inclusive jet EEC studied in~\cite{Apolinario:2025vtx,Chen:2024quk}, where the quark–gluon fraction changes significantly, altering the EEC shape due to quark and gluon jets having different shapes. Moreover, the normalization is unchanged within statistical uncertainties, indicating that the fraction of gluon jet energy carried by heavy-flavor hadrons is insensitive to modifications of the PDFs, as expected.

To quantify energy loss effects, we assume the quenching weight approximation~\cite{Baier:2001yt} to be valid, which directly extends~\eq{Fact} by a multiplicative convolution with partonic quenching factors $Q_i$. Specifically, we implement the simple prescription introduced in~\cite{Barata:2023bhh,Brewer:2025wol}, which accounts for both the fast thermalization of soft gluons and the out-of-cone energy carried by semi-hard emissions. We extend this construction from two-parton systems to jets with an arbitrary number of partons, as in~\cite{Brewer:2025wol}. All particles in a jet, at parton-level, are grouped into clusters of size $\theta_c = 2 / \sqrt{\hat{q} L^3}$, the characteristic scale for radiative decoherence. Each cluster is assigned a net flavor based on its parton content; cluster energy loss is then determined from the quenching weights by rescaling the momenta of each cluster by a factor $Q_i^{1/n}$, where $n=6$ is jet spectrum's spectral index. The 3-momentum of each particle is then rescaled by the cluster's energy loss weight. This prescription with the  EPPS16 nPDF set gives a good agreement with inclusive jet $R_{AA}$ at $L=4$ fm and $\hat{q}L=1.25\ \text{GeV}^2 / \text{fm}$ \cite{Brewer:2025wol}. Energy loss effects in the EEC are shown by the red curve in~\fig{EEC_eloss}, exhibiting the characteristic enhancement at small angles and suppression at large angles. This is due to the fact that jets prior to energy loss had higher $p_T$ values and thus the peak structure near $2p_T \sqrt{z} \sim m_Q$ in~\fig{EEC_vac} are shifted to the left relative to the one without energy loss. Unlike the inclusive jet case~\cite{Yang:2023dwc,Barata:2023vnl,Andres:2024pyz}, the overall normalization is not centered around unity. This again reflects the exclusive nature of the heavy-flavor EEC: the sum rule does not yield one when integrated over angles, in contrast to the inclusive EEC. The observed enhancement in normalization arises from the reduced energy loss of heavy quarks relative to light partons. We note that this enhancement is reduced when the EEC is normalized to the heavy-flavor jet cross section, since heavy-flavor
jets lose less energy than inclusive jets.

Finally, the blue curve in~\fig{EEC_eloss} shows the combined impact of energy loss, initial-state effects via nPDFs, and medium-modified splitting functions. The shape modification relative to the red curve, which has energy loss only but is similarly shaped even when nPDF is included, shows that adding the medium-modified splitting function modifies the shape much like it did relative to vacuum in~\fig{EEC_med}.

To mitigate the $p_T$ migration induced by energy loss, the quantile procedure developed in~\cite{Brewer:2018dfs,Apolinario:2024apr} can be applied. This exploits the approximate monotonicity of jet energy loss with respect to jet $p_T$: jets are ordered in $p_T$, and the $n^\text{th}$-highest-energy heavy-ion jet is matched to the $n^\text{th}$-highest-energy jet in vacuum from an equivalent sample. In this way, the comparison is made not at the same reconstructed $p_T$, but at the corresponding original jet $p_T$ before energy loss. Practically, we determine the quantile $p_T$ from the $c\bar{c}$-tagged jet sample and then apply it to select the inclusive jet sample.

The gold curve in~\fig{EEC_eloss} shows that the quantile procedure eliminates the small-angle enhancement shown by the red and blue curves, as the peak shift associated with reduced $p_T$ is removed. The resulting shape is much closer to the vacuum result in~\fig{EEC_med}, indicating that the procedure successfully corrects for kinematic migration. However, the overall normalization remains larger than in~\fig{EEC_med}. This reflects the fact that the quantile procedure does not remove energy loss itself, but only restores the appropriate vacuum reference for comparison. The enhanced normalization thus persists, consistent with the reduced energy loss of heavy quarks relative to light partons. Again, this enhancement is reduced when the EEC is normalized to the heavy-flavor jet cross section, which may help mitigate the effects of energy loss. An additional benefit of normalizing by the heavy-flavor-tagged jets is that it improves the monotonicity required for the quantile procedure since jets all have one flavor, though at the cost of sensitivity to the jet flavor definition.

\bibliography{EEC_ref.bib}{}
\bibliographystyle{apsrev4-1}
\newpage
\onecolumngrid
\newpage

\end{document}